\documentclass{llncs}

\usepackage{url}
\usepackage{listings}
\usepackage{owl}
\usepackage{tawny}
\usepackage{latex}
\usepackage{xspace}
\usepackage{alltt}
\usepackage{caption}
\usepackage{subcaption}
\usepackage{graphicx}

\usepackage[resetlabels]{multibib}
\newcites{iscn}{References}

\newcommand{\protege}{Prot\'eg\'e\xspace}
\newcommand{\latex}{\LaTeX\xspace} 

\lstMakeShortInline[style=kstyle]|

\lstnewenvironment{tawny}[1][]%
{\lstset{style=tawnystyle,frame=single,#1}}{}

\lstnewenvironment{latexlisting}[1][]%
{\lstset{style=latexstyle,frame=single,#1}}{}

\usepackage{karyotype}

\usepackage{listings}
\lstnewenvironment{code}[1][]%
{\lstset{style=kstyle,#1}}{}

\newcommand{\iscn}{ISCN\xspace}
\newcommand{\istring}{ISCN String\xspace}
\newcommand{\istrings}{ISCN Strings\xspace}
\newcommand{\ko}{The Karyotype Ontology\xspace}
\newcommand{\turners}{Tuners Syndrome\xspace}

\begin{document}

\title{A Highly Literate Approach to Ontology Building}

\author{Phillip Lord and Jennifer D. Warrender}

\institute{School of Computing Science, Newcastle
  University, Newcastle-upon-Tyne, UK}

\maketitle

\begin{abstract}
  Ontologies present an attractive technology for describing bio-medicine,
  because they can be shared, and have rich computational properties. However,
  they lack the rich expressivity of English and fit poorly with the current
  scientific ``publish or perish'' model. While, there have been attempts to
  combine free text and ontologies, most of these perform \textit{post-hoc}
  annotation of text. In this paper, we introduce our new environment which
  borrows from literate programming, to allow an author to co-develop both
  text and ontological description. We are currently using this environment to
  document the Karyotype Ontology which allows rich descriptions of the
  chromosomal complement in humans. We explore some of the advantages and
  difficulties of this form of ontology development.
\end{abstract}

\section{Introduction}
\label{sec:introduction}

Ontologies have been used extensively to describe many parts of bio-medicine.
Ontologies have two key features which make their usage attractive. First,
they provide a mechanism for standardizing and sharing the terms used in
descriptions, making comparison easier and, secondly, they provide a
computationally amenable semantics to these descriptions, making it possible
to draw conclusions about the relationships between descriptions even when
they share no terms in common.

Despite these advantages, the oldest and most common form of description in
biology is free text. Free text has numerous advantages compared to
ontologies: it is richly expressive, is widely supported by tooling, and while
the form of language used in science (``Bad English''~\cite{Wood_2001})
may not be easy to use, understand or learn, it is widely taught and most
scientists are familiar with it.

Between these two extremes of computable amenability, there are a full array
of different techniques. A ``database'' such as UniProt, for instance, appears
to be highly structured but also contains a large quantity of ``annotation''
that appears to be free text; although, even this contains informal structure,
which can be found and analysed by text analysis~\cite{michael}. We can set
this against descriptions of biological methods which appear in the form of a
scientific paper. The two forms of description have largely been used
independently. Ontology terms are used in semi-structured formats such as a
UniProt record, or minimum information documents, but in general, ontology
terms and the free text are in different parts of the record.

In this paper, we show how we can integrate ontological and textual
knowledge in a single authoring environment, and describe how we are
applying this to describing karyotypes.

\section{Developing Knowledge}
\label{sec:knowledge}

First, we ask the question, why is it difficult to relate ontological and
textual descriptions during authoring. One possible explanation is that the
two forms have very different ``development environments''\footnote{We lack a
  good term which covers word-processor, editor and IDE.}. The main
documentation environment used within science is Word, followed by \latex,
common in more mathematical environments. More recently, there has also been
interest in various light-weight markup languages, such as markdown. In the
case of Word, the development environment is a single tool which (effectively)
defines the file format, and the user interface that the author uses to
interact with it; with both \latex and other markup languages, there is a tool
chain in use, often with several options at each step, meaning that different
authors have (somewhat) different environments.

Ontology development environments also come in many different forms. Early
versions of the Gene Ontology, for instance, used a bespoke text file format
and a text editor -- an approach rather similar to the light-weight markup
languages of today. This had the significant advantage of a low-technological
barrier to entry, at least for authors, as well as easy integration with tools
such as version control systems which enabled collaborative working. It works
poorly using XML native formats like OWL (Ontology Web Language), however.
More modern environments, such as \protege and OBO-Edit provide a much more
graphical interface. These generally provide a much richer way of interacting
with an ontology; authors can see whole terms at once, using a variety of
syntaxes and allow rapid navigation through the class hierarchy, something
which most ontology authors do a lot~\cite{vigo2014protege4us}.

While these environments add a lot of value, they do not necessarily
integrate well with text. Both \protege and OBO-Edit have a
class-centric view and are biased toward showing the various logical
entities in the ontology, as opposed to the textual aspects. Indeed,
this bias is shown even at the level of OWL. For example, annotations
on an entity (or rather an axiom) are a \textit{set} rather than a
list, while ordering is generally considered to be essential for
most documents.

While there have been many attempts to integrate textual and
ontological knowledge, these have mostly involved \emph{post-hoc}
annotation of ontological entities using text analysis. A notable
exception to this is the Ontology Word add-in which uses text-analysis
to suggest ontology terms that can be used to annotate text at the
point of authorship~\cite{Fink_2010}.

\begin{figure*}[t]
  \centering
  \begin{subfigure}[t]{0.55\textwidth}
    \centering
    \includegraphics[width=\textwidth]{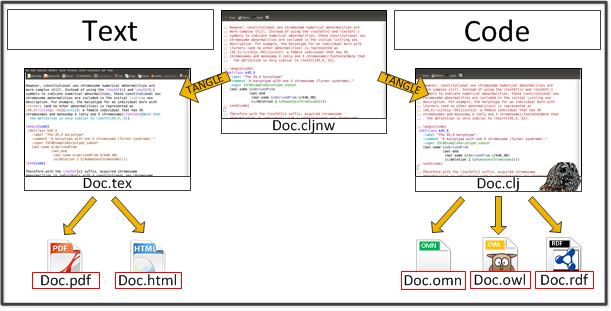}
    \caption{The traditional workflow}
    \label{fig:traditional}
  \end{subfigure}
  \begin{subfigure}[t]{0.55\textwidth}
    \centering
    \includegraphics[width=\textwidth]{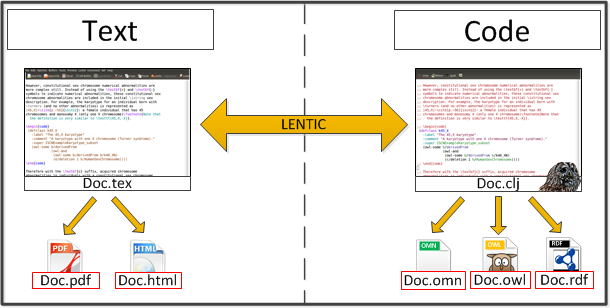}
    \caption{The lenticular workflow}
    \label{fig:lentic}
  \end{subfigure}
  \caption{Two workflows for the creation of literate ontologies. In a) the
    ontology is developed in an intermediate format (such as NoWeb), and the
    documentation and code-centric versions are generated. In b) both versions
    are developed simultaneously as views.}
\end{figure*}

With this divergence of development environments, it seems hard to
understand how we could square the circle of combining text and
ontology development. Next, we describe the Karyotype Ontology and how
the novel development methodology we used for this ontology allows us
to achieve this.

\section{The Karyotype Ontology}
\label{sec:karyotype-ontology}

A karyotype describes the number of chromosomes and any alterations from the
normal. These are visible under the light microscope, and when stained have a
characteristic banding pattern which can be used to distinguish between
different chromosomes and the positions on these chromosomes. In humans, these
alterations are described by their type, such as inversions, deletions or
duplications and by their location, specified by a chromosome number and band
number, following the \textit{ISCN} specification. So,
|46,XY,t(1;3)(p22;q13.1)| describes a male with a translocation from
chromosome |1p22| to chromosome |3q13.1|. The Karyotype Ontology is,
effectively, an ontological implementation of this ISCN specification for human
karyotype nomenclature~\cite{iscn09}\footnote{ISCN 2013 is now available}.

The Karyotype Ontology~\cite{warrender-karyotype} was a challenging
ontology to build because it is large but highly repetitive. It
provided the original motivation for and has been developed with
Tawny-OWL~\cite{tawny}, our novel ontology environment which provides
a fully programmatic development.  Tawny-OWL is implemented as a
Domain-Specific Language (or DSL) using the commodity Clojure language
and inherits its programmatic capabilities directly from there. Simple
ontological statements can be written with a syntax inspired by OWL
Manchester notation~\cite{ms2}; repetitive statements can be built
automatically by writing functions which encapsulate and abstract over
the simpler statements, a process we call
``patternisation''~\cite{warrender-pattern}. Tawny-OWL can be used to
generate any ontology and is not specific to the Karyotype Ontology;
the latter however is our most extreme example of a patternised
ontology with over 1000 classes using a single pattern. Tawny-OWL thus
fulfils the requirement for efficient population of an ontology,
something which tools like \protege are lacking~\cite{Vigo_2014}.

In addition to its programmatic capabilities, Tawny-OWL also allows the user
to take advantage of commodity Clojure development environments. For instance,
auto-complete, syntax highlighting, indentation and the REPL
(Read-Eval-Print-Loop, essentially a shell) all comes direct from Clojure. In
this way, we have managed to combine the advantages of text-based environments
for editing ontologies i.e. the use of a standard editing environment and
integration with version control, while maintaining (and in some ways
surpassing) the power of tools like \protege.

We consider next the implications that this has for the ability to
integrate ontological and textual descriptions.

\begin{figure*}[t]
\begin{subfigure}[t]{0.5\textwidth}
\begin{latexlisting}[basicstyle=\tiny\ttfamily]
In \ko, each karyotype is modelled by explicitly
stating the base karyotype and any abnormality
events, using the |b/derivedFrom| and
|e/hasDirectEvent| relations respectively. For this
exemplar, the base karyotype is |k/46,XX|, as the
tumour originated from a female. In addition, we
model the |1| deletion abnormality using a
cardinality restriction and the |e/Deletion| and
|h/HumanChromosome22| classes.

\begin{code}
(defclass k45_XX_-22
 :label "The 45,XX,-22 karyotype"
 :comment "A karyotype with monosomy 22."
 :super ISCNExampleKaryotype_subset
 (owl-some b/derivedFrom b/k46_XX)
 (exactly 1 e/hasDirectEvent
          (owl-and e/Deletion
                   h/HumanChromosome22)))
\end{code}
\end{latexlisting}
\caption{A document-centric view}
\end{subfigure}
\begin{subfigure}[t]{0.5\textwidth}

\begin{tawny}[basicstyle=\tiny\ttfamily]
;; In \ko, each karyotype is modelled by explicitly
;; stating the base karyotype and any abnormality
;; events, using the |b/derivedFrom| and
;; |e/hasDirectEvent| relations respectively. For this
;; exemplar, the base karyotype is |k/46,XX|, as the
;; tumour originated from a female. In addition, we
;; model the |1| deletion abnormality using a
;; cardinality restriction and the |e/Deletion| and
;; |h/HumanChromosome22| classes.

;; \begin{code}
(defclass k45_XX_-22
 :label "The 45,XX,-22 karyotype"
 :comment "A karyotype with monosomy 22."
 :super ISCNExampleKaryotype_subset
 (owl-some b/derivedFrom b/k46_XX)
 (exactly 1 e/hasDirectEvent
          (owl-and e/Deletion
                   h/HumanChromosome22)))
;; \end{code}
\end{tawny}
\caption{The ontology-centric view}
\end{subfigure}
\caption{Two lenticular views over an ontology source}
\label{fig:lentic-views}
\end{figure*}

\section{The Genesis of Literate Ontology}
\label{sec:literate-ontology}

As Tawny-OWL is based on a full programming language, it supports a feature
which at first seems quite inconsequential: comments. As with almost every
programming language, it is possible to add free, unstructured text to the
same source code that defines the ontology. While opinions vary on the role of
comments in programmatic code, perhaps the most extreme is that of literate
programming~\cite{lprogramming} which suggests that code should be usable
both as a program capable of execution and as a document capable of reading. 

A key aspect of literate programming is that neither view
should have primacy, which separates it from much weaker systems such as, for
example, JavaDoc, where the documentation very much fits into the code. We
call this form of development \textit{code-centric}. A more traditional
approach uses \textit{tangling}\footnote{The term ``tangling'' is not ours and
  is to our mind backward. However, it reflects the idea that source code is
  for consumption by a programmer and that this form is, therefore, untangled.
  The tangling process manipulates this clear form so that the computer can read
  it} -- here a single source document contains both ontological and document
source is created. It is then tangled to produce two forms of generated code
which in turn compile into the executable and documentation form (see
Figure~\ref{fig:traditional}). This form of editing is used by a number of
different systems, two of the most heavily used of which are DocTeX which
uses \latex to document \latex\footnote{Which is genuinely as confusing as it
  sounds} or Sweave~\cite{sweave} which combines \latex and
R~\cite{r-software}, the statistical programming language.

Our early attempts at literate ontology development used this approach. We
tried embedding OWL into \latex~\cite{literate-owl}. As an alternative, we
also build a system which allowed easy insertion of cross-references between a
\latex file and Manchester OWL notation~\cite{literate-omn}. However, we found
both to be highly-unusable. In one sense, tangling achieves the task of
putting the executable and documentable sections of a code-base on an equal
footing. However, in practice, there is a problem; the programmer has to edit
the untangled form. These days programmers are used to extremely rich
development environments which must be fully aware of the computational
amenable nature of the source code to function. In both cases, our early
experiments allowed the use of a \latex development environment, but provided
a very weak ontology development environment similar to the early use of text
editors. We call this form of development \textit{document-centric}. We found
this form of document-centric development so unattractive that it has been
abandoned.

\section{Literate Programming with Lenticular Views}
\label{sec:liter-progr-with}

The development of Tawny-OWL would make a tangling approach more viable, but
still we must choose: a document-centric approach would involve editing
Clojure source code without any IDE support (e.g. code evaluation, completion,
as well as indentation or syntax highlighting for the Clojure sections) while
a code-centric approach would lack support for \latex editing (e.g. citation
insertion, cross-referencing as well as indentation or syntax-highlighting for
the \latex sections).

Our latest solution attempts to square this circle. We provide a multi-view
approach to editing, which allows the author to see her source code in either
a \textit{document-centric} or a \textit{code-centric} view. We call this
approach \textit{lenticular} text, named after lenticular printing which
produces images which change depending on your angle of viewing. This is an
entirely novel approach to literate programming, effectively performing the
tangling operation for the author as they type. A representation of the two
views are shown in Figure~\ref{fig:lentic-views}. The two views, it should be
noted, contain the same \textbf{text} but are syntactically different, such
that the document-centric view is entirely valid \latex code, while the
ontology-centric view is valid Tawny-OWL code.

We have now implemented lenticular text for the editor,
Emacs\footnote{\url{https://www.gnu.org/software/emacs/}}, in a package called
``lentic''\footnote{\url{https://github.com/phillord/lentic}}. We choose Emacs
because it already provides a strong environment for editing both \latex and
Clojure\footnote{It also relatively easy to extend, and has support for
  Manchester OWL Notation added by one of us (PL).}
A key feature of this implementation is that both views exist
simultaneously in Emacs, and provide all the features of the appropriate
development environment; for example, ``tab-completion'' works in both the
document-centric view (completing \latex macros) and in the ontology-centric
view (completing ontology identifiers). We can launch a compilation of the
document-centric view (producing a PDF), or evaluate our ontology, perhaps
reasoning over it, in the code-centric view. Therefore, we have achieved a key
aim of literate programming: neither view holds primacy and the author can
edit either. The overall workflow is shown in Figure~\ref{fig:lentic}.

\section{A Literate Karyotype}
\label{sec:literate-karyotype}

The ISCN which describes karyotypes is an informal specification, combined with
many descriptions of particular karyotypes. For example, here we quote two
examples from page 56, ISCN 2009. These examples help to define the
specification further.

\begin{itemize}
\item \textbf{45,X} A karyotype with one X chromosome (Turner
  syndrome).
\item \textbf{47,XYY} A karyotype with one X chromosome and two Y
  chromosomes (Klinefelter syndrome).
\end{itemize}

In the Karyotype Ontology, we have encoded many of these examples, partly to
test that our ontology is capable of representing the ISCN specification.
Through the use of lenticular text, we are able to annotate these descriptions
both with references to the original work in ISCN as well as implementation
notes, describing our representation. We are steadily converting the whole of
the Karyotype Ontology into literate form; as an example of how this process
works, we have included the output of part of the Karyotype Ontology at the
end of this paper (see Section \ref{sec:appendix}). In short, the karyotype
ontology is becoming a fully literate ontology.

\section{Discussion}
\label{sec:discussion}

In this paper, we have described our methodology for integration of
text and ontological statements at authoring time, using lenticular
text to enable literate ontology development. This is a significant
advance over, for example, the Word Ontology plug-in, which enables
the use of ontology \emph{annotation} at authoring time. With
lenticular text, we are not limited to annotation with existing terms;
we can define terms of arbitrary complexity, allowing us to
post-coordinate our definitions~\cite{greycite24474}.

The combination of Tawny-OWL and lenticular text is an extremely rich
environment. We are aware, however, that it is a specialist environment. To
develop a literate ontology the author needs: to use Tawny-OWL, program in
Clojure, a Clojure development environment, write documents in \latex, and use
lentic package which is Emacs-based. In reality, though, the tools described
here are not tightly coupled. In particular:

\begin{itemize}
\item Clojure programming is only needed to extend Tawny-OWL.
\item Clojure is not tied to Emacs; there are other, well-supported
  environments.
\item Currently, lenticular text is novel and only implemented by the Emacs
  lentic package but it could be implemented in other
  environments\footnote{The first simple, version of lentic was around 1k loc,
    so this is not challenging to implement. Later versions are larger, as
    making the implementation efficient and scalable is somewhat harder.}
\item It is possible to edit a literate ontology \textit{without} using
  lenticular views, effectively replicating the traditional tangling workflow
  (see Figure~\ref{fig:traditional})\footnote{We actually use Lentic and Emacs in
    ``batch'' for this purpose, but an independent tool could be implemented
    very easily}.
\item Neither lenticular text nor the lentic package is specific to \latex or
  Tawny-OWL \footnote{Currently, lentic supports various combinations of Emacs-
    Lisp, Haskell or Clojure, with asciidoc, org-mode or \latex.}.
\item Both lenticular text and the lentic package are useful for general
  purpose programming and are not ontology specific\footnote{Lentic is
    self-documenting using Emacs-Lisp and org-mode, and Tawny-OWL is being
    converted. We also have entirely non-ontological users}.
\item Other embedded DSLs for OWL exist, such as
  ScOWL\footnote{\url{https://github.com/phenoscape/scowl}} and
  OWLJS\footnote{\url{https://github.com/cmungall/owljs}}.
\end{itemize}

While, we accept that the adoption of all the tooling described in the paper
maybe be relevant to very few developers, the use of parts of it have much
more widespread utility. It is, of course, unlikely to overtake Word as the
main tool for scientific authoring, it does have the potential to fulfil a
distinct niche as Sweave has done for statisticians.

We have, however, hit some problems with this process. We would
like to have developed the Karyotype Ontology alongside the text from
ISCN, so that the justification for each of the statements we have
made would be clear.  Unfortunately, the ISCN is published under a
non-permissive licence which prevents the production of this sort of
derived work. It is not even possible to hyperlink through to the
relevant sections of ISCN, as it is released only on paper. The irony
of our attempt to use Semantic Web technology on a resource that has
not even reached the web has not escaped our notice.

Likewise, our use of \latex integrates poorly with the web. While it is
possible to turn \latex source into HTML, it is not straight-forward. Lentic
supports other formats which are more suitable for this purpose (org-mode and
asciidoc) although they are formats aimed a programmers and have, for example,
comparatively weaker support for literate referencing. We also currently have
little support for cross-referencing \emph{between} the forms -- so
referring to ontology terms in text, for example, or sections in the
documentation from within ontology |rdfs:comment| annotations. We believe that
these extensions are entirely achievable in future.

Still, there are many other potential biomedical uses\footnote{As well as
  outside biomedicine: perhaps inenvitably, we have also used it to describe
  pizza.} for this form of technology, beyond karyotype descriptions. We are
currently also investigating clinical guidelines which describe treatment
plans -- fortunately in the UK, these are published with a permissive license.
In these cases, the knowledge being reproduced is such high value and
expensive to produce that the costs imposed by adding semantics in a
specialist environment are probably worthwhile. With Tawny-OWL and lentic, we
now have tools available which allow us to achieve this goal.

\section*{Acknowledgements}

This work was supported by Newcastle University.

\bibliographystyle{splncs03}
\bibliography{bibliography,phil_lord_refs,kblog}

\newpage
\begin{appendix}
\renewcommand\thesection{\Alph{section}}
\begin{sloppypar}





\section{Appendix: What is an \istring?}
\label{sec:appendix}

\setcounter{footnote}{0}

This section\footnote{This section is a demonstration of the
  output from our lenticular representation of
  karyotypes. It should not be considered to be a formal
  part of the paper.} provides a lenticular review of how
\istrings are defined by the specification and are modelled
using \ko, by focusing on a subset of exemplars defined in
the \iscn.

\begin{code}
;; Define namespace
(ns ^{:doc "Defining example karyotypes from the ISCN2013."
      :author "Jennifer Warrender"}
  ncl.karyotype.iscnexamples_subset
  (:use [tawny.owl])
  (:require [ncl.karyotype
             [karyotype :as k]
             [human :as h]
             [events :as e]
             [base :as b]]))

;; Define ontology
(defontology iscnexamples_subset
  :iri
  "http://www.purl.org/captau/karyotype/iscnexamples_subset"
  :prefix "iexs:"
  :comment "Subset of the ISCN Example Karyotypes ontology
  for Human Karyotype Ontology, written using the Tanwy_OWL
  library.")

;; Import all karyotype axioms
(owl-import k/karyotype)

;; Create a new subclass of Karyotype
(defclass ISCNExampleKaryotype_subset
  :super k/Karyotype)
\end{code}

In \ko ``normal'' karyotypes for each ploidy level are
modelled in the |base| ontology; thus we import all
associated axioms into the current ontology.

\begin{code}
(owl-import b/base)
\end{code}

However, not all karyotypes are normal; they can include a
variety of abnormalities. There are two types of
abnormality. \emph{Numerical abnormalities} are
abnormalities that affect the number of chromosomes present
in the karyotype, either by gaining or losing whole
chromosomes. \emph{Structural abnormalities} are
abnormalities that involve only parts of the
chromosomes\footnote{For simplicity, structural
  abnormalities will not be discussed at this time.}.

In order to model karyotypes, we need concepts in the
ontology that model the human chromosomes and the numerical
abnormality events. These are modelled in the |human| and
|events| ontologies respectively; thus we import all axioms
from both.

\begin{code}
(owl-import e/events)
(owl-import h/human)
\end{code}

In the \iscn, numerical abnormalities are represented in the
\istring using symbols and abbreviated terms. For numerical
abnormalities, the symbol \textbf{-} is used to represent
the loss of chromosomes while \textbf{+} represents the gain
of chromosomes.

For example, the karyotype of a female individual that has
lost one chromosome |22| (and no other abnormalities) is
represented as |k45,XX,-22|~\citeiscn[p.~57]{iscn12}; this
results in 45 chromosomes and monosomy (one copy of)
chromosome |22|.

In \ko, each karyotype is modelled by explicitly stating the
base karyotype and any abnormality events, using the
|b/derivedFrom| and |e/hasDirectEvent| relations
respectively. For this exemplar, the base karyotype is
|k/46,XX|, as the tumour originated from a female. In
addition, we model the |1| deletion abnormality using a
cardinality restriction and the |e/Deletion| and
|h/HumanChromosome22| classes. However due to the
programmatic nature of Tawny-OWL, we can implement
parameterised patterns~\citeiscn{warrender-pattern}, thus
simplifying the deletion abnormality definition to one line
of code, using the |e/deletion| pattern.

\begin{code}
(defclass k45_XX_-22
  :label "The 45,XX,-22 karyotype"
  :comment "A karyotype with monosomy 22."
  :super ISCNExampleKaryotype_subset
  (owl-some b/derivedFrom b/k46_XX)
  (e/deletion 1 h/HumanChromosome22))
\end{code}

Similarly, the karyotype of a tumour from a female
individual that has lost one chromosome |X| (and no other
abnormalities) is represented as
|k45,X,-X|~\citeiscn[p.~56]{iscn12}. In \ko, this karyotype is
modelled with the base karyotype |b/46,XX| and |1| deletion
event that involves |h/HumanChromosomeX|.

\begin{code}
(defclass k45_X_-X
  :label "The 45,X,-X karyotype"
  :comment "A tumor karyotype in a female with loss of one X
  chromosome."
  :super ISCNExampleKaryotype_subset
  (owl-some b/derivedFrom b/k46_XX)
  (e/deletion 1 h/HumanChromosomeX))
\end{code}

However, the classification of abnormalities is not so
simple; an abnormality can be also classified as either a
\textit{constitutional} or \textit{acquired}
abnormality\footnote{All previous exemplars define acquired
  abnormalities.}. A constitutional abnormality, also known
as an in-born abnormality, is an abnormality that is present
in (almost) all cells of an individual and exists at the
earliest stages of embryogenesis, while an acquired
abnormality is an abnormality that develops in somatic
cells~\citeiscn{constitutional}.

Generally, constitutional abnormalities are indicated using
the suffix \textbf{c}. For example the \istring
|46,XY,+21c,-21|~\citeiscn[p.~58]{iscn12} represents the
karyotype of tumour cells taken from a male individual, that
had a constitutional trisomy |21| and has acquired disomy
|21|. Using this representation we see that karyotypes with
constitutional abnormalities explicitly define two types of
canonicalisation; one of the individual and the other for
the cell line they have given rise to.

In \ko, constitutional abnormalities are also modelled
explicitly using the |e/hasDirectEvent| relation. However
unlike acquired abnormalities, constitutional abnormalities
are modelled as a nested restriction in conjunction with the
base karyotype. In this exemplar:
\begin{itemize}
\item the base karyotype is |b/46,XY| (as the karyotype
  originates from a male individual).
\item the |1| constitutional abnormality is a gain of one
  chromosome |21|. The associated parameterised pattern for
  gain is |e/addition|.
\item the |1| acquired abnormality is a loss of one
  chromosome |21|.
\end{itemize}

\begin{code}
(defclass k46_XY_+21c_-21
  :label "The 46,XY,+21c,-21 karyotype"
  :comment "Acquired loss of one chromosome 21 in a patient
  with Down syndrome."
  :super ISCNExampleKaryotype_subset
  ;;aka 47,XY,+21
  (owl-some b/derivedFrom
           (owl-and
            (owl-some b/derivedFrom b/k46_XY)
            (e/addition 1 h/HumanChromosome21)))
  (e/deletion 1 h/HumanChromosome21))
\end{code}

However, constitutional sex chromosome numerical
abnormalities are more complex still. Instead of using the
\textbf{+} and \textbf{-} symbols to indicate numerical
abnormalities, these constitutional sex chromosome
abnormalities are included in the initial \istring sex
description. For example, the karyotype for an individual
born with \turners (and no other abnormalities) is
represented as |45,X|~\citeiscn[p.~56]{iscn12}: a female
individual that has 45 chromosomes and monosomy X (only one
X chromosome)\footnote{Note that the definition is very
  similar to \texttt{45,X,-X}.}.

\begin{code}
 (defclass k45_X
   :label "The 45,X karyotype"
   :comment "A karyotype with one X chromosome (Turner
   syndrome)."
   :super ISCNExampleKaryotype_subset
   (owl-some b/derivedFrom
             (owl-and
              (owl-some b/derivedFrom b/k46_XN)
              (e/deletion 1 h/HumanSexChromosome))))
\end{code}

With the \textbf{c} suffix, acquired chromosome
abnormalities in individuals with a constitutional sex
chromosome abnormality can easily be distinguished. For
example the \istring |46,Xc,+21|~\citeiscn[p.~57]{iscn12}
represents tumour cells taken from a female individual with
\turners; a constitutional monosomy X and an acquired
trisomy 21.

\begin{code}
(defclass k46_Xc_+21
  :label "The 46,Xc,+21 karyotype"
  :comment "Tumor cells with an acquired extra chromosome 21
  in a patient with Turner syndrome."
  :super ISCNExampleKaryotype_subset
  ;;aka 45,X
  (owl-some b/derivedFrom
            (owl-and
             (owl-some b/derivedFrom b/k46_XN)
             (e/deletion 1 h/HumanSexChromosome)))
  (e/addition 1 h/HumanChromosome21))
\end{code}

\begin{code}
;; Implement disjoint axioms
(as-disjoint k45_XX_-22 k45_X_-X
             k46_XY_+21c_-21 k45_X k46_Xc_+21)
\end{code}

Now that we defined a few exemplar karyotypes, we discuss
the definition of sex.

\subsection{Defining Sex}
\label{sec:sex}

While building this ontology, we found that sex is not as
intuitive as it seems. The obvious definition for sex was
that a ``male'' karyotype should be defined as a karyotype
with a |Y| chromosome, while a ``female'' karyotype as one
without. However further investigation showed that these
definitions are, in fact, too simplistic as the karyotype
|45,X,-Y|\footnote{A male-derived cell line which has lost
  its \texttt{Y} chromosome.}, has no |Y| chromosome, yet
would generally be considered to be a ``male'' karyotype.

Therefore, the finalised definition for sex, as shown below
considers the history of the karyotype by asserting a
|derivedFrom| relation\footnote{Due to the transitive
  property of \texttt{b/derivedFrom}, we can also determine
  the sex of karyotypes that contain constitutional
  abnormalities.}. Using these definitions, the |45,X,-Y|
karyotype can be correctly stated as being a ``male''
karyotype.

\begin{code}
(defclass MaleKaryotype
  :equivalent
  (owl-or
   b/k46_XY
   (owl-some b/derivedFrom b/k46_XY)))
\end{code}

\begin{code}
(defclass FemaleKaryotype
  :equivalent
  (owl-or
   b/k46_XX
   (owl-some b/derivedFrom b/k46_XX)))
\end{code}

However these definitions are unable to ontologically
categorise the |45,X| karyotype as either female or male
though it would generally be considered a ``female''
karyotype. There is no correct answer to this problem. We
could either redefine our female karyotype to include the
|45,X| karyotype or add phenotypic sex. This decision needs
to be taken by the domain experts themselves.

\bibliographystyleiscn{abbrv}
\bibliographyiscn{bibliography}


\end{sloppypar}
\end{appendix}

\end{document}